\documentstyle [12pt,epsf] {article}

\catcode`@=12
\def\bra#1{\left\langle #1\right|}
\def\ket#1{\left| #1\right\rangle}
\def\be{\begin{equation}}
\def\ee{\end{equation}}
\def\lsim{\mathrel{\vcenter{\hbox{$<$}\nointerlineskip\hbox{$\sim$}}}}
\newcommand{\bea}{\begin{eqnarray}}
\newcommand{\eea}{\end{eqnarray}}
\newcommand{\nn}{\nonumber}

\def\sss{\scriptscriptstyle}
\begin{document}
\vspace{0.5in}
\oddsidemargin -.375in
\newcount\sectionnumber
\sectionnumber=0
\def\bra#1{\left\langle #1\right|}
\def\ket#1{\left| #1\right\rangle}
\def\be{\begin{equation}}
\def\ee{\end{equation}}
\def\lsim{\mathrel{\vcenter{\hbox{$<$}\nointerlineskip\hbox{$\sim$}}}}
\thispagestyle{empty}


\def\sss{\scriptscriptstyle}
\def\barp{{\raise.35ex\hbox
{${\sss (}$}}---{\raise.35ex\hbox{${\sss )}$}}}
\def\barpd{{\raise.35ex\hbox
{${\sss (}$}}--{\raise.35ex\hbox{${\sss )}$}}}
\def\dbbarp{\hbox{$B^{0}$\kern-1.2em\raise1.5ex\hbox{\barpd}}}
\def\kbarp{\hbox{$K^{*0}$\kern-1.6em\raise1.5ex\hbox{\barpd}}}

\def\dkbarp{\hbox{$K^{*}$\kern-1.2em\raise1.5ex\hbox{\barpd}}}
\vskip0.5truecm

\def\tr{\mbox{tr}\,}
\def\Tr{\mbox{Tr}\,}
\def\dag{^\dagger}
\def\res{\mbox{Res}}
\def\re{\mbox{Re}\,}
\def\b{\bigskip}
\def\s{\smallskip}
\def\l{\hspace*{0.05cm}}
\def\esp{\hspace*{1cm}}

\def\be{\begin{equation}}
\def\ee{\end{equation}}
\def\bea{\begin{eqnarray}}
\def\eea{\end{eqnarray}}

\begin{center}

{\large \bf
\centerline{Direct CP violation and new physics effects}} 
{\large \bf
\centerline
{in the decay mode {\boldmath{$B^+ \rightarrow \phi K^+$}}}}

\vspace*{1.0cm}
{ Anjan K. Giri$^1$ and  Rukmani Mohanta$^2$ } \vskip0.3cm
{\it  $^1$ Physics Department, Technion-Israel 
Institute of Technology, 32000 Haifa, Israel}\\
{\it $^2$ School of Physics, University of Hyderabad,
Hyderabad - 500046, India
} \\
\vskip0.5cm
\bigskip
(\today)
\vskip0.5cm

\begin{abstract}
We study the direct CP violation effect in the decay mode $B^+ \to 
\phi K^+ $.
This decay mode is dominated by the loop induced $\bar b \to \bar s s
\bar s  $ 
penguin diagram with a tiny contribution from the annihilation  diagram.
Therefore, the standard model expectation of direct CP violation is 
negligibly small. 
Using QCD factorization approach we find the CP asymmetry in the standard
model to be at percent level. We consider then
two scenarios beyond the standard model, 
the model with an extra vector-like down quark (VLDQ) and
the R-parity violating supersymmetric model (RPV)
 and show that the 
direct CP violating
asymmetry in $B^+ \to \phi K^+ $  could be as large as $\sim 85 \% $
($ 70 \% $) in VLDQ (RPV) model.
\end{abstract}
\end{center}
\thispagestyle{empty}
\newpage
\baselineskip=14pt

\section{Introduction}

The study of $B$ physics and CP violation is now at the center stage of 
high energy physics research with dedicated $B$ factories (BABAR and Belle)
already having an accumulation of huge data in the $B$ sector. 
With these two $B$ factories in full operation and hadronic 
machines coming up, the flavor sector of the standard model (SM) 
will be subjected to stringent tests in the near future. 
The major goal of these $B$ factories is not only to test the predictions of 
the SM but also to reveal the presence of new physics (NP), if any. 
CP violation in $B$ system has been confirmed recently at both the
$B$ factories with the measurement of $\sin 2\beta$ (where $\beta $
is one of the angles of CKM unitarity triangle) in the $B^0
\to \psi K_S$ decay \cite{belle0}. It should be emphasized 
here that the current measured value of $(\sin 2 \beta)_{\psi K_S} $ 
by both the $B$
factories are very close to each other \cite{tom1}. 
The present world avarage on $(\sin 2 \beta)_{\psi K_S} $ \cite{tom1} is

\be
\sin(2 \beta)_{\psi K_S}=0.736 \pm 0.049\;,
\ee
which is in very good agreement with the SM expectation. 
However, this result does 
not exclude intersting CP violating new physics effects in other $B$ decays.
The rare decay mode $B^0 \to \phi K_S$, which is a pure penguin process,
involving the quark level transition $b \to s \bar s s $, is one of the
channels which provides powerful testing ground for new physics. 
The reason is very simple because in the SM the direct decay amplitudes
for  $B^0 \to \psi K_S$ and  $B^0 \to \phi K_S$ modes have vanishing weak phase
(in the Wolfenstein parametrization). Thus, the time dependent
mixing-induced CP asymmetry in both decays is due to the weak phase
in $B^0 - \bar B^0$ mixing and expected to give the same value, i.e., 
$\sin 2 \beta$. In the SM, the difference between these two measurements 
is expected to be very small 
\cite{gross1}   
\be
|\sin(2 \beta)_{\psi K_S}-\sin(2 \beta)_{\phi K_S}| \lsim {\cal O} 
(\lambda^2)\;,
\ee
with $\lambda \sim 0.22$. However, the recent 
measurements of sin 2$\beta$ in the $B^0 \to \phi K_S$ decay 
at BABAR and Belle 
give the average value \cite{tom1} 
\be
\sin(2 \beta)_{\phi K_S}=-0.15 \pm 0.33\;,
\ee 
which is about $2.7 \sigma $ deviation away 
from the corresponding measurement of $(\sin 2\beta)_{\psi K_S}$.
The discrepancy between the measured values of $(\sin 2\beta)_{\psi K_S}$
and $(\sin 2\beta)_{\phi K_S}$ may imply the possible existence of 
new physics in one of the decay modes.

If there is new physics, then it could affect the $B^0 - \bar B^0$
mixing as well as the decay amplitudes. But the NP contribution
arising from mixing, will be same (i.e., universal)
in both the processes .
So when we compare
the measurements of $\sin 2\beta$ in both the two decays then the NP 
contribution, if any in mixing, will not appear in the comparison, 
since it will affect both the decays with the same amount. 
Whereas, the NP contribution
present in the decay amplitudes are nonuniversal and process dependent
and can vary from process to process. The striking difference between these 
two decays is that $B^0 \to \psi K_S$  is a tree level 
($b \to c \bar c s$)  process whereas $B^0 \to \phi K_S$ is a 
purely loop induced penguin process. So the NP contribution to
$B^0 \to \psi K_S$ is expected to be suppressed and it could be significant
for the loop induced process $B^0 \to \phi K_S$. So, the discrepancy between 
the measured values of $(\sin 2\beta)_{\psi K_S}$ and
$(\sin 2\beta)_{\phi K_S}$ may indicate the possibility of NP effect 
in the decay amplitude of $B^0 \to \phi K_S$.

Recently, various new physics scenarios have been explored to explain the 
above discrepancy \cite{new1,datta1,hiller1,rm1}. 
If new physics effects indeed are present in 
the decay mode $B^0 \to \phi K_S$, then one can expect to observe similar 
effects in other modes having the same internal quark structure.
It is therefore important to explore other signals of new physics in order to
corroborate this result. One way to search for new physics effects is
to look for direct CP violation  in the decay modes which are having a single
decay amplitude in the standard model and hence expected to give zero 
direct CP asymmetry.
An observation of nonzero direct CP violation 
in such modes is an unambiguous signal of new physics. 

At present, it appears from the current trend of data that new physics 
effects might be present in the $B^0 \to \phi K_S$ mode.
It is therefore interesting to see the effects of NP in the 
penguin dominated decay mode
$B^\pm \to \phi K^{\pm}$ also (which is having the same underlying 
quark dynamics) alongwith a tiny annihilation
contribution,
which is our prime objective in this paper.
The branching ratio and CP asymmetry measurements for this mode have  
recently been reported by both the Belle \cite{bel3}, BABAR \cite{bab3}
and CLEO \cite{cleo} 
collaborations, which are given below as
\bea
{\rm Br}(B^+\to \phi K^+) & =& (9.4 \pm
 1.1 \pm 0.7)\times 10^{-6}~~~~~~~~~
{\rm Belle}\;, \nn\\
{\cal A}_{\rm CP}(B^+ \to \phi K^+)& =& 0.01 \pm 0.12 \pm 0.05
~~~~~~~~~~~~~~~~{\rm Belle}\;,\nn\\
{\rm Br}(B^+ \to \phi K^+) & = & (10.0^{+0.9}_{-0.8} \pm 0.5)\times 10^{-6}
~~~~~~~~~~{\rm BABAR}\;,\nn\\
{\cal A}_{\rm CP}(B^+ \to \phi K^+) & = & 0.04 \pm 0.09 \pm 0.01
~~~~~~~~~~~~~~~~{\rm BABAR}\;,\nn\\
{\rm Br}(B^+ \to \phi K^+) & = & (5.5^{+2.1}_{-1.8} \pm 0.6)\times 10^{-6}
~~~~~~~~~~~~{\rm CLEO}\;.
\label{eq:data}
\eea
The average branching ratio and CP asymmetry are given as
\bea
{\rm Br}(B^+ \to \phi K^+) & = & (9.2 \pm 0.7)\times 10^{-6}\;,\nn\\
{\cal A}_{\rm CP}(B^+ \to \phi K^+) & = & 0.03 \pm 0.07 \;.\label{expt}
\eea
At this point we ask the question whether the possibility of NP effect in 
the decay mode  $B^+ \to \phi K^+ $ has already been ruled out ?
In the rest of the paper
our objective will be to closely examine this question and try to obtain a
meaningful answer, if any.
If one looks at the data on the direct CP asymmetry,
then certainly nothing can be concluded at present. 
That is the central 
value is higher than the SM expectation, but as 
error bars are also significantly large, one simply cannot
conclude/exclude the possibility of NP effect. So the present 
status is that, it is premature to say anything and to be more precise,
on which side (i.e., within or outside) of the SM they are, as  far as 
direct CP violation in $B^\pm \to \phi K^\pm $ is concerned.
To summarize, although the CP violating asymmetry is not very much 
larger than the SM expected value, but due to the presence of large error bars,
decisive conclusion regarding the presence/absence of NP effect cannot 
be inferred.

This in turn gives us enough room to explore NP effect. In fact if in future 
the data stabilize with even a few percent of direct CP asymmetry,  then it 
may be very hard to explain the same under the framework of the SM and 
eventually that may lead to the establishment of NP in this mode. 
Keeping this in mind we now study 
carefully to find the answer to our question that we raised earlier.

In this paper, we intend to study the direct CP violation effects in the decay
mode $B^\pm \to \phi K^\pm $. We use the QCD factorization method 
\cite{beneke1,beneke2} to
evaluate the relevant branching ratio and the direct
CP asymmetry ${\cal A}_{\rm CP}$, in the SM. 
This decay mode $B^\pm \to \phi K^\pm $ has recently
been studied in the SM using QCD factorization \cite{he1,cheng1,du1}.
However, the predicted branching ratio found to be 
well below the present experimental value.
Next, we consider two beyond the standard model scenarios, 
the so called R-parity violating supersymmetric 
model and the model with an extra vector-like down quark.
These two models were explored recently to explain the observed 
discrepancy in $B^0 \to \phi K_S $ mode \cite{datta1, hiller1, rm1}.
 
The paper is organized as follows.
Section II includes a general description of CP violating parameter
in $B^\pm \to \phi K^\pm $ decays, while in Sect. III we analyze the
decay mode $ B^\pm \to  \phi K^{\pm}$ in the SM using the QCD factorization. 
The new physics effects from the VLDQ model and the RPV model are considered in
sections IV and V respectively and in Section \ VI we present 
some concluding remarks.

\section{CP violating Asymmetry}

Here we briefly present the basic and well known formula for the CP violating
asymmetry parameter. For charged $ B^\pm \to \phi K^\pm$ decays 
the CP violating rate asymmetries in the partial rates are defined as follows :
\be
{\cal A}_{\rm CP}= \frac{\Gamma(B^- \to \phi K^-)-\Gamma(B^+ \to \phi K^+)}
{\Gamma(B^- \to \phi K^-)+\Gamma(B^+ \to \phi K^+)}\;.\label{cp0}
\ee
Without loss of generality we can write the decay amplitudes as
\bea
A(B^- \to \phi K^-) &=& \lambda_u |A_u| e^{i \delta_u}
+\lambda_c |A_c| e^{i \delta_c}\;,\nn\\
A(B^+ \to \phi K^+) &=& \lambda_u^* |A_u| e^{i \delta_u}
+\lambda_c^* |A_c| e^{i \delta_c}\;,\label{dcp}
\eea
where $\lambda_q= V_{qb}V_{qs}^*$, $A_u$ and $A_c$ denote
the contributions arising from the current operators
proportional to the product of CKM matrix elements $\lambda_u$
and $\lambda_c$ respectively. The corresponding strong phases are denoted by
$\delta_u$ and $\delta_c$. Thus the CP violating
asymmetry is given as
\bea
{\cal A}_{\rm CP} &=& \frac{-2 {\rm Im}(\lambda_u \lambda_c^*)
{\rm Im}(A_u A_c^*)} 
{|\lambda_u A_u|^2+|\lambda_c A_c|^2 + 2{\rm Re}(\lambda_u \lambda_c^*)
{\rm Re}( A_u A_c^*)}\nn\\
&=& \frac{2 \sin \gamma \sin (\delta_u -\delta_c)}
{|\frac{\lambda_u A_u}{\lambda_c A_c}| +|\frac{\lambda_c A_c}{\lambda_u A_u}|
+2 \cos \gamma \cos (\delta_u -\delta_c)}\;,\label{dcp1}
\eea
where $-\gamma$ is the weak phase of $\lambda_u$, and $\lambda_c$ is real in
the Wolfenstein parametrization. Thus to obtain significant 
direct CP asymmetry, one requires the two interfering
amplitudes to be of same order and their relative strong phase should be
significantly large (i.e., close to $\pi/2$). However, in the SM, the ratio 
of the CKM matrix elements of the two terms in Eq. (\ref{dcp}) 
can be given (in the Wolfenstein
parametrization) as $|\lambda_u/\lambda_c|\simeq \lambda^2 \sqrt{\rho^2+\eta^2}
\simeq 2 \% $. Therefore, the first amplitude will be highly suppressed 
with respect to the second unless $A_u >> A_c$.   
Therefore, the naive  
SM expectation on ${\cal A}_{\rm CP}$ is that it is negligibly small.
This in turn makes the mode interesting to look for the NP in terms of
large direct CP asymmetry.   

In the presence of new physics the amplitude can be written as
\be
A(B^- \to \phi K^-)=A_{SM}+A_{NP}= A_{SM}\left [ 1+ r_{NP}~ e^{i
\phi_{NP}} \right ]\;,\label{br11}
\ee
where $r_{NP}=|A_{NP}/A_{SM}|$, ($A_{SM}$
and $A_{NP}$ correspond to the SM and NP contributions to the $B^-
\to \phi K^-$ decay amplitude respectively) and $\phi_{NP}={\rm
arg}(A_{NP}/A_{SM} )$, which contains both strong and weak phase
components.
The branching ratio for the $B^- \to \phi K^-$ decay process can be given as
\be
{\rm Br}(B^- \to \phi K^-)={\rm Br^{SM}}\left ( 1+r_{NP}^2 +2 r_{NP} \cos
\phi_{NP} \right )\;,
\ee
where ${\rm Br^{SM}}$ represents the corresponding standard model value.

To find out the CP asymmetry, it is necessary to represent explicitly 
the strong and
weak phases of the SM as well as NP amplitudes. Although, it is expected that
the SM amplitude $\lambda_u A_u$ is highly suppressed with 
respect to its
$\lambda_c A_c$ counterpart, for completeness we will keep this term
for the evaluation of ${\cal A}_{\rm CP}$. We denote the NP contribution
to the decay amplitude as $A_{NP}=|A_{NP}| e^{i(\delta_n+\theta_n)}$,
where $\delta_n$ and $\theta_n$ denote the strong and weak phases of the NP
amplitude respectively. Thus, in the presence of NP, we can explicitly 
write the decay amplitude for
$B^- \to \phi K^-$ mode as
\be
A(B^- \to \phi K^-) =\lambda_u |A_u| e^{i \delta_u}
+\lambda_c |A_c| e^{i \delta_c}+|A_{NP}| e^{i(\delta_n+\theta_n)}
\;.\label{al}
\ee
The amplitude for $B^+ \to \phi K^+$ mode is obtained by changing the
sign of the weak phases of the amplitude (\ref{al}). Thus,
the CP asymmetry parameter (\ref{cp0}) is given as
\be
{\cal A}_{\rm CP} =\frac{2\biggr(|\lambda_u \lambda_c A_u A_c|
\sin \gamma \sin \delta_{uc}+|\lambda_u A_u A_{NP}|
\sin (\gamma+\theta_n) \sin \delta_{un}+
| \lambda_c A_c A_{NP}|
\sin \theta_n \sin \delta_{cn}\biggr)}
{|{\cal A}|^2+
2\biggr\{|\lambda_u A_u|\biggr(|\lambda_c  A_c|
\cos \gamma \cos \delta_{uc}+| A_{NP}|
\cos (\gamma+\theta_n) \cos \delta_{un}\biggr)+
| \lambda_c A_c A_{NP}|
\cos \theta_n \cos \delta_{cn}\biggr\}},
\ee
where $|{\cal A}|^2=(|\lambda_u A_u|^2+|\lambda_c A_c|^2+|A_{NP}|^2)$
and $\delta_{ij}=\delta_i-\delta_j $ are the relative strong phases
between different amplitudes. If we neglect the  $\lambda_u A_u$ 
component in the decay amplitude (\ref{al}),
the CP asymmetry will be reduced to
\be
{\cal A}_{\rm CP} =\frac{2\biggr(
| \lambda_c A_c A_{NP}|
\sin \theta_n \sin (\delta_{c}-\delta_n)\biggr)}
{|\lambda_c A_c|^2+|A_{NP}|^2+
2\biggr(
| \lambda_c A_c A_{NP}|
\cos \theta_n \cos (\delta_c-\delta_n)\biggr)}.\label{acp1}
\ee
Having obtained the direct CP asymmetry parameter ${\cal A}_{\rm CP}$ in 
the presence of new physics we now proceed to evaluate the same 
within and beyond the SM, in the
following sections.

\section{CP Violation in $B^\pm \to \phi K^\pm $
process in the SM} 

To study the CP violation effects in $B^\pm \to \phi K^\pm$
process, first we present the SM amplitude and find out the branching ratio
and CP asymmetry parameter. 
In the SM, the decay process $B^\pm \to \phi K^\pm$ receives dominant
contribution from the
quark level transition $b \to s \bar s s$, which is induced by the
QCD, electroweak and magnetic penguins and a tiny annihilation 
contribution. The effective $\Delta B= \Delta S =1 $ 
Hamiltonian relevant for the process under consideration is
given as \cite{beneke1} 
\be 
H_{eff}=
\frac{G_F}{\sqrt{2}}\biggr[V_{ub}V_{us}^*\sum_{i=1}^2 C_i O_i+
V_{qb}V_{qs}^* \sum_{j=3}^{11}C_j O_j
 \biggr],
\ee
where $q=u,~c$. $O_{1,2}$, $O_3, \cdots, O_{6}$ and $O_7,
\cdots, O_{10}$ are the standard model tree, QCD and EW penguin operators,
respectively, and $O_{11}$ is the gluonic magnetic operator.
The values of the Wilson coefficients at the scale $\mu \approx m_b$
in the NDR scheme are given
in Ref. \cite{Buc96} as
\bea
&&C_1=1.082\;,~~~~C_2=-0.185\;,~~~~C_3=0.014\;,~~~~~~C_4=-0.035\;,\nn\\
&&C_5=0.009\;,~~~~C_6=-0.041\;,~~~~C_7=-0.002\alpha\;,~~~C_8=0.054 \alpha\;,
 \nn\\
&&C_9=-1.292\alpha\;,~~~~C_{10}=0.263 \alpha\;,~~~~C_{11}=0.052\;.
\eea

We use QCD factorization \cite{beneke1, beneke2}  
to evaluate the hadronic matrix elements.
In this method, the decay amplitude can be represented in the form
\be
\langle \phi K^- |O_i |B^- \rangle =
\langle \phi K^- |O_i |B^- \rangle_{\rm fact}\biggr[
1+\sum r_n \alpha_s^n + {\cal O}(\Lambda_{\rm QCD}/m_b)
\biggr]\;,
\ee
where $\langle \phi K^- |O_i |B^- \rangle_{\rm fact}$ denotes the 
naive factorization result and $\Lambda_{\rm QCD} \sim 225$ MeV, 
the strong interaction scale. The second and third terms in the square
bracket represent higher order $\alpha_s$ and $\Lambda_{\rm QCD}/m_b$
corrections to hadronic matrix elements. 
We shall closely follow the approaches 
\cite{he1,cheng1,du1,yang1} to reanalyze our QCD factorization
calculation.
In a recent paper Beneke and Neubert have 
presented the latest QCD factorization calculation
\cite{beneke3}, where the authors have considered 4 different schemes 
to find a suitable set of parameters which can explain the observed data. 
Furthermore, it has been shown in that paper that the last scheme,
i.e., S4 (scheme 4) explains the observed data on branching ratios and CP 
violation parameters to a good accuracy. Of course it should be mentioned 
here that even the scheme S4 also fails to explain the problematic penguin 
dominated $B \to \phi K$ modes \cite{beneke3, neubert1}.   

In this analysis, however, we will restrict ourselves to the default QCD 
factorization (by default we mean the 
QCD factorization result without any parameter readjustment (for more details
the reader is referred to Ref. \cite{beneke3}).
This is done so, because of the fact that one should not just ignore the 
possibility of NP effect. Furthermore, we do not see why the parameter 
readjustment is the only possibility
to match the data (eventhough it may be a logistic option and in the long 
run may even come out to be true).
So, in order to visualize the NP effect, we confine ourselves to the default 
values of the QCD factorization and allow NP to fill up the gap between 
the experimental
data and the SM values rather than readjusting the parameters to find a good 
agreement.  In doing so, we can accommodate the beyond the SM
scenarios, if any, to explain the possible discrepancy. 
Nevertheless, we will compare
the results of Ref. \cite{beneke3} wherever necessary to that of our work.   
It should be emphasized here that our speculation regarding the presence of
NP may not be unrealistic. As a 
matter of fact the present data in this sector do not seem to comply with the
SM expectations, already mentioned earlier, 
as in the case of $B^0 \to \phi K_S$ decay.

In the heavy quark limit the decay amplitude for the $B^- \to \phi K^- $
process, arising from the penguin diagrams, is given  as 
\bea
A^P(B^- \to \phi K^-) = \frac{G_F}{\sqrt 2}\sum_{q=u,~c} 
V_{qb}V_{qs}^* \biggr[a_3^q+a_4^q+a_5^q-\frac{1}{2}\left (
a_7^q+a_9^q+a_{10}^q \right )+a_{10a}^q \biggr]X\;,
\eea
where $X$ is the factorized matrix element. Using the form factors and decay
constants defined as \cite{bsw}
\bea
\langle K^-(p_K) |\bar s \gamma ^\mu b | B^-(p_B) \rangle &=&
\biggr[(p_B+p_K)^\mu-\frac{m_B^2-m_K^2} {q^2} q^\mu\biggr] F_1(q^2)\nn\\
&+&
\frac{m_B^2-m_K^2}{q^2}q^\mu F_0(q^2)\;,\nn\\
\langle \phi(q, \epsilon) |\bar s \gamma ^\mu s | 0 \rangle &=&
f_\phi~ m_\phi~ \epsilon^\mu\;,
\eea
we obtain 
\bea
 X &=& \langle  K^- (p_K)| \bar s
\gamma_\mu(1-\gamma_5)b | B^-(p_B) \rangle
\langle \phi(q, \epsilon )|\bar s
\gamma^\mu(1-\gamma_5)s|0 \rangle \nn\\
& = & 2 F^{B \to K}_1(m_\phi^2)~f_\phi~ m_\phi~
(\epsilon \cdot p_B)\;.
\eea
The coefficients $a_i^q$'s which contain next to leading order
(NLO) and hard scattering corrections are given as \cite{he1, yang1}
\bea
a_3^u &=& a_3^c ~=~ C_3+\frac{C_4}{N}+\frac{\alpha_s}{4 \pi}\frac{C_F}{N}
C_4 F_\phi \;,\nn\\
a_4^q &= & C_4+\frac{C_3}{N}+\frac{\alpha_s}{4 \pi} \frac{C_F}{N}
\biggr[ C_3 \left [F_\phi + G_\phi(s_s) + G_\phi(s_b) \right ]\;,
\nn\\
&+&C_1 G_\phi(s_q) +(C_4+C_6) \sum_{f=u}^b G_\phi(s_f)
+C_{11} G_{\phi, 11}\biggr]\;,\nn\\
a_5^u &=& a_5^c ~=~ C_5+\frac{C_6}{N}+\frac{\alpha_s}{4 \pi}
\frac{C_F}{N}C_6(- F_\phi -12) \;,\nn\\
a_7^u &=& a_7^c ~=~ C_7+\frac{C_8}{N}+\frac{\alpha_s}{4 \pi}
\frac{C_F}{N}C_8(- F_\phi -12) \;,\nn\\
a_9^u &=& a_9^c ~=~ C_9+\frac{C_{10}}{N}+\frac{\alpha_s}{4 \pi}
\frac{C_F}{N}C_{10} F_\phi \;,\nn\\
a_{10}^u &=& a_{10}^c ~=~ C_{10}+\frac{C_9}{N}+\frac{\alpha_s}{4 \pi}
\frac{C_F}{N}C_9 F_\phi \;, \nn\\
a_{10a}^q &= & \frac{\alpha_s}{4 \pi} \frac{C_F}{N}
\biggr[( C_8+C_{10}) \frac{3}{2}
\sum_{f=u}^{b}e_f G_\phi(s_f)\nn\\
& + & C_9 \frac{3}{2}\left (e_s G_\phi(s_s) +e_b G_\phi(s_b) \right )
\biggr]\;,\label{qcd}
\eea
where $q$ takes the values $u$ and $c$, $N=3$, is the number of
colors, $C_F=(N^2-1)/2N$. The internal quark mass in the
penguin diagrams enters as  $s_f=m_f^2/m_b^2$. 
The other parameters in (\ref{qcd}) are given as
\bea
F_\phi &=& -12 \ln \frac{\mu}{m_b}-18+f_\phi^I +f_\phi^{II}\;,\nn\\
f_\phi^I &= & \int_0^1 dx~ g(x) \phi_\phi(x)\;,\nn\\
g(x) &= & 3 \frac{1-2x}{1-x}\ln x -3i \pi\;, \nn\\
f_\phi^{II} &= & \frac{4 \pi^2}{N} \frac{f_K f_B}{F_1^{B \to K}(0) m_B^2}
\int_0^1 \frac{dz}{z} \phi_B(z) \int_0^1 \frac{dx}{x} \phi_K(x) 
\int_0^1 \frac{dy}{y} \phi_\phi(y)\;, \nn\\
G_\phi(s) &=& \frac{2}{3}-\frac{4}{3}{\rm ln}\frac{\mu}{m_b}
+4\int_0^1 dx~ \phi_\phi(x) \nn\\
&\times & \int_0^1 du~u (1-u) \ln\left [ s-u(1-u)(1-x) -i \epsilon 
\right ]\;,
\nn\\
G_{\phi,11} &=& - \int_0^1 dx \frac{2}{1-x}\phi_\phi(x)\;.
\eea
The light cone distribution amplitudes (LCDA's) at twist two order 
are given as
\bea
&&\phi_B(x) = N_B x^2(1-x)^2 {\rm exp}\biggr(-\frac{m_B^2 x^2}{2
\omega_B^2}\biggr) \;,\nn\\
&&\phi_{K, \phi}(x)=  6x(1-x)\;,
\eea
where $N_B$ is the normalization factor satisfying
$ \int_0^1 dx \phi_B(x)=1$ and $\omega_B=0.4$ GeV. The quark masses appear in
$G(s)$ are pole masses and we have used the following values (in GeV)
in our analysis;
\begin{eqnarray*}
m_u=m_d=m_s=0, ~~~~m_c=1.4~~~~m_b=4.8.
\end{eqnarray*}
The contributions arising from the  annihilation diagrams are
given \cite{cheng1,du1} as
\bea
A^A(B^- \to \phi K^-) & = &\frac{G_F}{\sqrt 2} f_B f_K f_\phi \frac{2 m_\phi}
{m_B^2}(\epsilon \cdot p_B)\biggr[
V_{ub}V_{us}^*  \left \{ b_2(\phi, K^-)+b_3(\phi, K^-)
+b_3^{ew}(\phi, K^-) \right \}\nn\\
&+ &V_{cb}V_{cs}^*  \left \{ b_3(\phi, K^-)
+b_3^{ew}(\phi, K^-) \right \}\biggr]
\eea
The parameters $b_i$ \cite{du1} are as 
\bea
b_2(\phi, K^-) &=& \frac{C_F}{N^2} C_2 A_1^i(\phi, K^-) \;,\nn\\
b_3(\phi, K^-) &=& \frac{C_F}{N^2} \biggr [
C_3 A_1^i(\phi, K^-) +C_5 A_3^i(\phi, K^-)
+\left ( C_5+N C_6 \right ) A_3^f(\phi, K^-)\biggr]\;,\nn\\
b_3^{ew}(\phi, K^-) &=& \frac{C_F}{N^2} \biggr [
C_9 A_1^i(\phi, K^-) +C_7 A_3^i(\phi, K^-)
+\left ( C_7+N C_8 \right ) A_3^f(\phi, K)^-\biggr]\;,
\eea
where
\bea
A_1^i(\phi, K^-) &=& \pi \alpha_s \int_0^1 dx~ \phi_\phi(x)\int_0^1 dy~
\phi_K(y) \times \left [ \frac{1}{y(1-x(1-y))} + \frac{1}{y(1-x)^2} 
\right] \;,\nn\\
A_3^i(\phi, K^-) &=& \pi \alpha_s \int_0^1 dx~ \phi_\phi(x)\int_0^1 dy~
\phi_K(y) \times r_\chi  \frac{2(1-y)}{y(1-x)(1-x(1-y))}\;, 
\nn\\
A_3^f(\phi, K^-) &=& \pi \alpha_s \int_0^1 dx~ \phi_\phi(x)\int_0^1 dy~
\phi_K(y) \times r_\chi  \frac{2(2-x)}{y(1-x)^2}\;. 
\eea
These integrals contain the divergent end point integrals. Assuming
SU(3) flavor symmetry and symmetric LCDA's (under $x \leftrightarrow
(1-x)$), the weak annihilation contributions can be parametrized as
\bea
A_1^i(\phi, K^-) &\simeq & 18 \pi \alpha_s \left [ X_A -4 +\frac{\pi^2}
{3} \right]\;,\nn\\
A_3^i(\phi, K^-) &\simeq & \pi \alpha_s r_\chi \left [ 2\pi^2
+6(X_A^2 -2X_A) \right]\;,\nn\\
A_3^f(\phi, K^-) &\simeq & 6 \pi \alpha_s r_\chi \left [2 X_A^2 
-X_A \right ]\;,
\eea 
where $X_A=\int_0^1dx/x$ parameterizes the divergent end point integral and
$r_\chi=2 m_K^2/m_b(m_s+m_u)$ denotes the chiral enhancement factor.
It should be noted that the quark masses in the chiral enhancement factor
are running quark masses and we have used their values  at the $b$ 
quark mass scale
 as $m_b(m_b)$=4.4 GeV, $m_s(m_b)$=90 MeV and $m_u(m_b)=3.2$ MeV.
Thus we obtain the total amplitude as (in units of $10^{-4}$)
\bea
A(B^- \to \phi K^-)&=& -\frac{G_F}{\sqrt 2}~(\epsilon \cdot p_B)~\biggr[
V_{ub}V_{us}^*\left (45.022 + i~47.788 \right )+V_{cb}V_{cs}^*
\left (51.407+i~
28.054 \right )\biggr]\nn\\
&=& -\frac{G_F}{\sqrt 2}~(\epsilon \cdot p_B)~\biggr[
V_{ub}V_{us}^* \left (65.656~ e^{i47^\circ}
\right )+V_{cb}V_{cs}^*\left (58.564~ e^{i29^\circ}\right )
\biggr]\;.
\label{kl1}
\eea
The branching ratio can be obtained using the formula
\bea
{\rm Br}(B^- \to \phi K^-) = \tau_{B^-} \frac{|p_{\rm cm}|^3}{
8 \pi m_\phi^2}~|{A(B^- \to \phi K^-)}/({\epsilon \cdot p_B})|^2\;,
\eea
where $p_{\rm cm}$ is the momentum of the outgoing particles in the $B$
meson rest frame. 

We have used the following input parameters. The value of the form factor 
at zero recoil is taken as $F_1^{B \to K}(0)=$ 0.38, 
and its value at $q^2=m_\phi^2$ can be obtained 
using simple pole dominance ansatz \cite{bsw} as 
$F_1^{B \to K}(m_\phi^2)=$ 0.39. 
The  values of the decay constants are as $f_{\phi}=$ 0.233 GeV, 
$f_B=0.19 $ GeV,
$f_K$=0.16 GeV,   and the lifetime of $B^-$ meson $\tau_{B^-}=
1.674 \times 10^{-12}$ sec \cite{pdg}. For the CKM matrix elements,
we have used the 
Wolfenstein parameterization and have
taken the values of the parameters
$A=0.819$, $\lambda=0.2237$, $\rho=0.224$ and $\eta=0.324$.
With these input parameters,  we obtain
the branching ratio in the SM as
\be
{\rm Br^{SM}}(B^- \to \phi K^-)=5.99 
\times 10^{-6}\;,
\ee
which lies quite below the present experimental
limit (\ref{expt}).
The corresponding CP asymmetry parameter ${\cal A}_{\rm CP}$ is found to be
\be
{\cal A}_{\rm CP}=0.011\;,
\ee
which is also below the present central experimental value
(\ref{expt}). 
However, our predicted branching ratio is in close agreement 
with the various earlier 
calculations \cite{cheng1,du1} and with 
that of the default value in \cite{beneke3}. The small difference 
between the predicted values in these analyses is due to difference 
in the used input parameters (i.e quark masses, decay constants and 
CKM parameters etc.). It should be noted here that the obtained value 
of branching ratio is slightly
above the predicted value of Ref. \cite{he1}, where they have
not considered the annihilation contributions.  
The ${\cal A}_{\rm CP}$ value obtained here
is in agreement with the previous calculation of Beneke and Neubert
\cite{beneke3}. 
It should be pointed out here that even the best scheme S4 \cite{beneke3},
which predicts close to the branching ratio 
obtained by the experiments, predicts even smaller ${\cal A}_{\rm CP}$ (0.007).
Now if we consider the error bars in the data, certainly
upto 10 \% of  ${\cal A}_{\rm CP}$ is allowed by the present data. 
Such a large ${\cal A}_{\rm CP}$,
if established later,
certainly cannot be accommodated in the SM. Thus in turn this brings us to the 
door of NP to provide some meaningful explanation. It can be seen from
Eq. (\ref{kl1}) that since $|A_u| \simeq |A_c|$ and also 
their strong phases ($\delta_u$ and $\delta_c$) are 
nearly equal, the observed CP asymmetry is quite small. 
Furthermore, as discussed in section II,
since $|A_u| \simeq |A_c|$ and 
$V_{ub}V_{us}^*/V_{cb}V_{cs}^* \simeq 2\% $, we will neglect the
$\lambda_u A_u $ term in the decay amplitude (\ref{al}) 
when we consider the contributions from 
beyond standard model scenarios. 

\section{Contribution from the VLDQ Model}

Now we consider the model with an additional vector like down
quark \cite{ref11}. It is a simple model beyond the SM with an
enlarged matter sector with an additional vector-like down quark
$D_4$. The most interesting effects in this model concern CP
asymmetries in neutral $B$ decays into final CP eigenstates. At a
more phenomenological level, models with isosinglet quarks provide
the simplest self consistent framework to study deviations of $3
\times 3$ unitarity of the CKM matrix as well as allow flavor
changing neutral currents at the tree level. The presence of an
additional down quark implies a $4 \times 4$ matrix $V_{i \alpha}$
$(i=u,c,t,4,~\alpha= d,s,b,b')$, diagonalizing the down quark mass
matrix. For our purpose, the relevant information for the low
energy physics is encoded in the extended mixing matrix. The
charged currents are unchanged except that the $V_{CKM}$ is now the $3
\times 4$  upper submatix of $V$. However, the distinctive feature
of this model is that FCNC enters neutral current Lagrangian of
the left handed downquarks :
 \be
 {\cal L}_Z= \frac{g}{2 \cos
\theta_W} \left [ \bar u_{Li} \gamma^{\mu} u_{Li} - \bar d_{L
\alpha}U_{\alpha \beta} \gamma^\mu d_{L \beta}-2 \sin^2 \theta_W
J_{em}^\mu \right ]Z_{\mu}\;,
\ee
with
\be
U_{\alpha \beta}
=\sum_{i=u,c,t} V_{\alpha i}^\dagger V_{i \beta} =\delta_{\alpha
\beta} - V_{4 \alpha}^* V_{4 \beta}\;,
\ee
where $U$ is the neutral
current mixing matrix for the down sector which is given above. As
$V$ is not unitary, $U \neq {\bf{1}}$. In particular its
non-diagonal elements do not vanish :
\be
U_{\alpha \beta}= -V_{4
\alpha}^* V_{4 \beta} \neq 0~~~{\rm for}~ \alpha \neq \beta\;,
\ee

Since the various $U_{\alpha \beta}$ are non vanishing they would
signal new physics and the presence of FCNC at tree level, this
can substantially modify the predictions for CP asymmetries. The
new element $U_{s b}$ which is relevant to our study is given as
\be
U_{sb}= V_{us}^* V_{ub}+V_{cs}^*V_{cb}+V_{ts}^*V_{tb}\;.
\ee

The decay modes $B^\pm \to \phi K^\pm $ receive the new contributions 
both from 
color allowed and color suppressed $Z$-mediated FCNC transitions.
The new additional operators are given as
\bea
&&O_1^{Z-FCNC}=[\bar s_\alpha \gamma^\mu (1-\gamma_5) b_\alpha][\bar s_\beta
\gamma_\mu(C_V^s -C_A^s \gamma_5) s_\beta] \;,\nn\\
&&O_2^{Z-FCNC}=[\bar s_\beta \gamma^\mu (1-\gamma_5) b_\alpha][\bar s_\alpha
\gamma_\mu(C_V^s -C_A^s \gamma_5) s_\beta]\;,
\eea
where $C_V^s$ and $C_A^s$ are the vector and axial vector $Z s \bar s$
couplings. Using Fierz transformation and the identity
$(C_V^s-C_A^s \gamma_5)=[(C_V^s+C_A^s)(1- \gamma_5)+(C_V^s-C_A^s)(1+ \gamma_5)]
/2$, as done in \cite{rm1}, the  amplitude, including the
nonfactorizable contributions,  is 
given  as
\be 
A^{VLDQ}(B^- \to \phi K^-)=\frac{G_F}{\sqrt 2} U_{sb}~2X
\left (C_V^s+\frac{C_A^s}{3}\right )\left (1 +
\frac{\alpha_s}{4 \pi}\frac{C_F}{N} 
F_\phi\right ) \;.
\ee
The values for $C_V^s$ and $C_A^s$ are taken as
\be
C_V^s= -\frac{1}{2}+\frac{2}{3} \sin^2 \theta_W\;,
~~~~~~~~~C_A^s=-\frac{1}{2}\;.
\ee
Now using  $\sin^2 \theta_W$=0.23, alongwith
$U_{sb}=|U_{sb}| e^{i \theta_n}$, where $\theta_n$
is the new weak phase of VLDQ model and 
\be
|U_{sb}| \simeq 1 \times 10^{-3}\;,
\ee 
 we find the amplitude 
\be
A^{VLDQ}(B^- \to \phi K^-) =  -\frac{G_F}{\sqrt 2} e^{i \theta_n} ~
(\epsilon \cdot p_B)\left ( 1.683-i~0.140 \right ) 
\times 10^{-4}\;.
\ee
Thus we obtain the strong phase in VLDQ model to be
$\sim -5^\circ$ and $r_{NP}$ as
\be 
r_{NP}=0.7\label{np1}\;.
\ee
From Eq. (\ref{np1}), it should be noted that the NP amplitude is  
of the same order as the SM amplitude and hence large 
CP asymmetry is expected in this model.
Now plotting the  branching ratio (\ref{br11}) versus $\phi_{NP}$
(the red line in Figure-1),
we see that the observed data can be easily accommodated in   
VLDQ model. The direct CP asymmetry ${\cal A}_{\rm CP}$ (Eq. (\ref{acp1})
vs. $\theta_n$ is plotted in Figure-2, for $\delta_c=29^\circ$ and
$\delta_n=-5^\circ$, and it is also (the red line in Figure-2)
in compatible with the 
present experimental data. Furthermore, it should be noted
from the Figure-2 that
VLDQ model can accommodate the CP violating asymmetry upto
$85 \%$.  

\begin{figure}[htb]
   \centerline{\epsfysize 2.5 truein \epsfbox{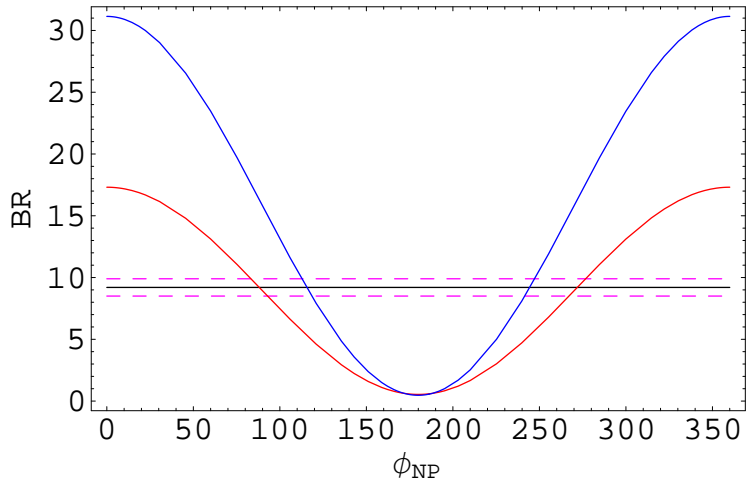}}
   \caption{
  Branching ratio of $B^- \to \phi K^{-}$ process 
 (in units of $10^{-6}$) versus the phase
 $\phi_{NP}$ (in degree). The red and blue curve represent 
the results of VLDQ and RPV model respectively.
 The horizontal solid line is the central experimental
 value whereas the dashed horizontal lines denote the error
 limits.}
  \end{figure}



\begin{figure}[htb]
   \centerline{\epsfysize 2.5 truein \epsfbox{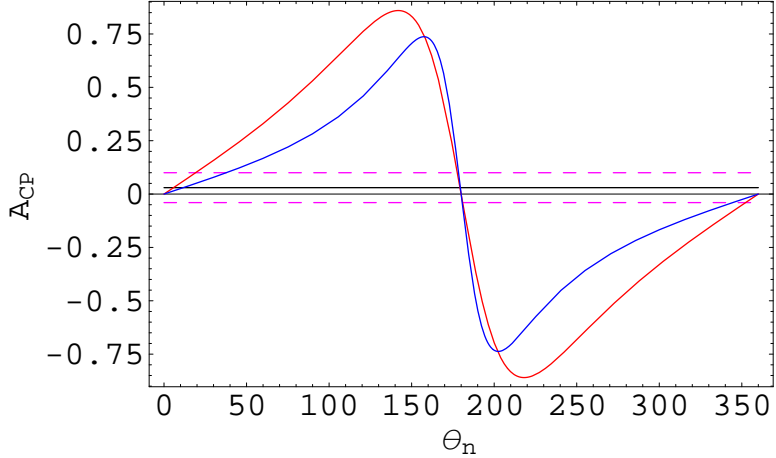}}
   \caption{
  The direct CP asymmetry (${\cal A}_{\rm CP}$) in the 
 $B^\pm \to \phi K^{\pm}$ process 
 versus the new weak phase
 $\theta_{n}$ (in degree). The red and blue curve represent 
the results of VLDQ and RPV model.
 The horizontal solid line is the central experimental
 value whereas the dashed horizontal lines denote the error
 limits.}
  \end{figure}



\section{Contribution from R-Parity violating supersymmetric model}

We now analyze the decay mode in the minimal supersymmetric model with
R-parity violation.
In the supersymmetric models there may be interactions which
violate the baryon number $B$ and the lepton number $L$
generically. The simultaneous presence of both $L$ and $B$ number
violating operators induce rapid proton decay, which may contradict
strict experimental bound. In order to keep the proton lifetime
within experimental limit, one needs to impose additional symmetry
beyond the SM gauge symmetry to force the unwanted baryon and lepton
number violating interactions to vanish. In most cases this has
been done by imposing a discrete symmetry called R-parity defined
as, $R_p=(-1)^{(3B+L+2S)}$, where $S$ is the intrinsic spin of the
particles. Thus the $R$-parity can be used to distinguish the
particle ($R_p$=+1) from its superpartner ($R_p=-1$). This
symmetry not only forbids rapid proton decay, it also renders
stable the lightest supersymmetric particle (LSP). However, this
symmetry is ad hoc in nature. There is no theoretical arguments in
support of this discrete symmetry. Hence it is interesting to see
the phenomenological consequences of the breaking of R-parity in
such a way that either $B$ and $L$ number is violated, both are
not simultaneously violated, thus avoiding rapid proton decay.
Extensive studies has been done to look for the direct as well as
indirect evidence of R-parity violation from different processes
and to put constraints on various R-parity violating couplings.
The most general $R$-parity and Lepton number violating
super-potential is given as
\begin{equation}
W_{\not\!{L}} =\frac{1}{2} \lambda_{ijk} L_i L_j E_k^c
+\lambda_{ijk}^\prime L_i Q_j D_k^c \;,\label{eq:eqn10}
\end{equation}
where, $i, j, k$ are generation indices, $L_i$ and $Q_j$ are
$SU(2)$ doublet lepton and quark superfields and $E_k^c$, $D_k^c$
are lepton and down type quark singlet superfields. Further,
$\lambda_{ijk}$ is antisymmetric under the interchange of the
first two generation indices. Thus the relevant four fermion
interaction induced by the R-parity and lepton number violating
model is \cite{gue1}
\bea
{\cal H}_{\not\!{R}}  &= &
\frac{1}{8 m^2_{\tilde \nu}}\biggr[
(\lambda_{i23}^{\prime *} \lambda_{i22}^{\prime})
(\bar s_\alpha \gamma^\mu(1-\gamma_5)s_\beta)~
(\bar s_\beta \gamma_\mu (1+\gamma_5)b_\alpha)
\nn\\
&& \hspace{ 0.2 cm}+
(\lambda_{i32}^{\prime} \lambda_{i22}^{\prime *})
(\bar s_\alpha \gamma^\mu(1+\gamma_5)s_\beta)~(\bar s_\beta 
\gamma_\mu (1-\gamma_5)b_\alpha)
\biggr]\;.\label{al1}
\eea
It should be noted that, the factorized matrix elements of both the 
operators in Eq. (\ref{al1}) are same because of
\bea
&&\langle K^-|\bar s \gamma^\mu(1-\gamma_5)b|B^- \rangle
=\langle K^-|\bar s \gamma^\mu(1+\gamma_5)b|B^- \rangle=
\langle K^-|\bar s \gamma^\mu b|B^- \rangle \;,\nn\\
&&\langle \phi|\bar s \gamma^\mu(1-\gamma_5)s|0 \rangle
=\langle \phi|\bar s \gamma^\mu(1+\gamma_5)s|0 \rangle =
\langle \phi|\bar s \gamma^\mu s|0 \rangle\;.
\eea
So, including the nonfactorizable effects, we obtain the R-parity 
violating contribution to the decay amplitude as
\bea
A^{\not\! R}(B^- \to \phi K^-) =
\frac{1}{8 m^2_{\tilde \nu}} 
\Big(\lambda_{i32}^{\prime } \lambda_{i22}^{\prime *}
+\lambda_{i23}^{\prime *} \lambda_{i22}^{\prime}\Big)X
 \biggr[ \frac{1}{N} +\frac{\alpha_s}{4 \pi}\frac{C_F}{N}
(-F_\phi -12)
\biggr]\;,
\eea
where the summation over $i=1,2,3$ is implied.
Now considering the values of R-parity couplings from \cite{datta1}
as
\be
-\frac{1}{ 24 m^2_{\tilde \nu}} 
\Big(\lambda_{i32}^{\prime } \lambda_{i22}^{\prime *}
+\lambda_{i23}^{\prime *} \lambda_{i22}^{\prime}\Big)=\frac{R}{12M^2}
e^{i \theta_n}\;,
\ee
where $\theta_n$ is the new weak phase of R-parity violating
couplings and 
 $R \sim 1.5 \times 10^{-3} $ for the mass scale 
$M \sim m_{\tilde \nu}=100$ GeV. 
 
Thus we obtain the R-parity violating amplitude as
\be
A^{RPV}(B \to \phi K^-)=-(\epsilon \cdot p_B)
e^{i \theta_n}\left (2.462+i~0.512 
\right )\times 10^{-9}\;,
\ee
with the strong phase as $\sim 12^\circ$. Thus, the
new physics contribution $r_{NP}$ is found to be
\be
r_{NP}=1.28\;.
\ee
Plotting the branching ratio (\ref{br11}) vs. $\phi_{NP}$ we can see that
the observed branching ratio can be easily accommodated in the RPV model
(blue curve in Fig-1). Similarly the observed direct 
CP asymmetry can also
be explained by RPV model, (blue curve in Figure-2). As can be seen 
from Fig-2, the RPV model can accommodate CP asymmetry upto 70$\%$.
 
\section{Conclusion}

The measured branching ratio and direct CP asymmetry parameter
in the  $B^\pm \to \phi K^{\pm} $ mode are not in
agreement with the SM expectations. The direct CP asymmetry in this mode in
the SM is expected to be very small (at $1\%$ level) but the measured
value is higher than this. But this 
should be treated with caution since the error bars are still very large in 
comparison to the central value and definitive conclusion cannot be obtained.
On the other hand there is a very wild speculation and/or a general belief 
that exist in the literature is that there might be some kind of new physics 
effect present,  in the decay amplitude
of the pure penguin process $B^0 \to \phi K_S $, to account
for the $2.7 \sigma$ deviation of the measured $(\sin 2 \beta)_{\phi K_S}$
from $(\sin 2 \beta)_{\psi K_S}$.
So it is quite obvious that presence of NP may not be ruled out 
on its charged counterpart process $B^\pm \to \phi K^{\pm} $ having the same
underlying quark dynamics.

In fact there are various ways to test the presence of new physics that exist
in the literature. One of them is to find nonzero direct CP asymmetry in 
modes which are dominated by a single decay amplitude in the SM and hence
expected to give zero direct CP asymmetry. The mode $B^\pm \to \phi K^\pm $
falls into that class, where the
${\cal A}_{\rm CP}$ is expected to be negligibly small.
But the present data do not seem to respect the 
same and may reveal
hidden NP effects in future. In order to look for NP, we first carefully 
reanalyzed this mode ($B^\pm \to \phi K^{\pm} $) in the default QCD 
factorization approach. We also took into account of the small annihilation 
contribution in our calculation
and found that the branching ratio and ${\cal A}_{\rm CP}$ are 
deviated from the present experimetal data. 
However, our results are in close agreement with the latest
theoretical calculation \cite{beneke3}.

We took this opportunity to incorporate the possible new physics scenarios to
explain the same. 
In order to obtain significant direct CP asymmetry both the interfering
amplitudes (i.e., the SM and NP amplitudes) should be of the 
same order and the relative phase differences between them 
should be nonvanishing.
We then introduce two beyond the standard model scenarios in turn 
to explain the observed branching ratio 
and ${\cal A}_{\rm CP}$. These are the VLDQ model and the 
RPV supersymmetric 
model. These two models are shown in the literature that they 
can successfully explain the possible discrepancy
of $B^0 \to \phi K_S$ result with that of the SM expectation. 
We find that $r_{NP}$, which is the ratio between the NP
and SM amplitudes, to be 0.7 and 1.28 for VLDQ and RPV model respectively.
As can be seen now from the  
Figures-1 and 2, that these two models can also accommodate the possible
deviation of branching ratio and CP asymmetry parameter 
from the SM values for the
decay under consideration. Eventually,
it turns out that direct CP asymmetry upto $\sim 85 \% ~ (70\%)$ can be 
accommodated in the VLDQ (RPV) model, if nature is so obliged 
to give such a large
value. In fact more precise 
data in future will lead us to definitive conclusions and/or narrow down some
NP scenarios.

To conclude, the branching ratio and direct CP asymmetry 
parameter (${\cal A}_{\rm CP}$)
as obtained in Belle and BABAR indicate deviation 
from that of the SM expectation in the
$B^\pm \to \phi K^\pm $ decay. If the trend remains the same, as it is now, 
then it may lead to the establishment of the presence of new physics 
in penguin dominated
($b\to s \bar s s $) $B$ decays. At present it is too early to say anything
in favor of or rule out the possibility of the 
existence of new physics in this sector.
At the same time keeping in mind all these we should keep every option open, 
explore various possibilities and hope that $B$ factory
data will reveal new physics in the near future. 

\section{Acknowledgments}
AKG was supported by the Lady Davis Fellowship and
the work of RM was supported in part by Department of Science and
Technology, Government of India through Grant No. SR/FTP/PS-50/2001.

\end{document}